# Physical-layer Network Coding: A Random Coding Error Exponent Perspective

Shakeel Salamat Ullah, Gianluigi Liva, and Soung Chang Liew

*Abstract*—In this work, we derive the random coding error exponent for the uplink phase of a two-way relay system where physical layer network coding (PNC) is employed. The error exponent is derived for the practical (yet sub-optimum) XOR channel decoding setting. We show that the random coding error exponent under optimum (i.e., maximum likelihood) PNC channel decoding can be achieved even under the sub-optimal XOR channel decoding. The derived achievability bounds provide us with valuable insight and can be used as a benchmark for the performance of practical channel-coded PNC systems employing low complexity decoders when finite-length codewords are used.

## I. INTRODUCTION

In a two-way relay channel (TWRC), two end nodes exchange their messages via a relay node. In TWRC operated with physical-layer network coding (PNC) [1], [2], the end users transmit simultaneously to the relay in the uplink phase. The relay receives the superimposed signal and aims to recover a linear combination (e.g., the bit-wise XOR when operating over a binary finite field) of the messages transmitted by the two end nodes. In the downlink phase, the relay broadcasts the linear combination back to both end nodes. The end nodes subtract their contribution to the linear combination from the received message to recover their intended message.

Different decoding strategies can be applied at the relay [3]–[5]. Optimum decoding (in the maximum likelihood sense) requires searching for the most likely linear combination. More specifically, a maximum-likelihood (ML) PNC channel decoder shall compute the likelihood for a given information message by finding all codeword pairs $(\mathbf{a}, \mathbf{b}) \in \mathcal{C}^{\text{A}} \times \mathcal{C}^{\text{B}}$ yielding the considered linear combination (assuming the same rate for both codes), where $\mathcal{C}^{\text{A}}$ and $\mathcal{C}^{\text{B}}$ are the codes used to transmit by the two end nodes. This approach entails a large complexity, and can be applied only to a few classes of error correcting codes [6], [7]. When the end users employ the same linear block code $\mathcal{C}$, a simpler (suboptimal) approach consists in deriving an overall likelihood for the symbol pairs that yield each linear combination. The symbol-wise likelihoods, one for each possible linear combination, are then used by the decoder to provide a decision on the linear combination of the overall codewords[1]. When binary codes are used, this approach is ofter referred to as XOR channel decoding (XOR-CD). In the remainder of the paper, we will focus on the binary case only.

In [8], [9] the random coding error exponent [10] for uplink PNC transmission was derived. The analysis considered an ML PNC channel decoder for a case where a common binary linear code was adopted at both end nodes. The error exponent derived can be used to obtain performance bounds for PNC systems; however, the analysis does not provide an in-depth understanding of practical PNC systems employing low-complexity channel decoding such as XOR-CD. The ML PNC channel decoder has prohibitive computational complexity, and it is not feasible for practical PNC implementations, even assuming finite-length codewords.

In this work, we derive the random coding error exponent for practical PNC systems employing XOR-CD. We show that the random coding error exponent derived in [8], [9] for a linear PNC system under ML decoding is achievable under XOR-CD as well despite the sub-optimality of XOR-CD. The derived achievability bounds provide us with key insights on the finite-length performance of a practical channel coded PNC systems and can be used as benchmark for XOR-CD schemes based on practical decoders. The results are obtained by introducing an equivalent degraded channel (EDC) model that simplifies the analysis of XOR-CD.

The rest of the paper is organized as follows. Section II presents the system model. The equivalent degraded channel model is introduced in Section III. Section IV applies the EDC model to the calculation of random coding error exponent under both perfect channel-state-information (CSI) and the mismatched CSI setting. Section V concludes the paper.

## II. SYSTEM MODEL

We consider a PNC system under a TWRC setting. Two end nodes (labelled as A and B in the following) exchange their information via a relay node in two time slots only. In the first time slot, the two users transmit simultaneously to the relay. Given the noisy observation of the users' superimposed signal, the relay performs decoding according to the XOR-CD setting. The recovered message is then encoded and broadcasted to both users in the second time slot. The two end nodes then recover their intended messages by subtracting their own messages from the network-coded message.

This work is supported by .... grant .... and the General Research Funds Project Number ...., established under the University Grant Committee of the Hong Kong Special Administrative Region, China. This work is also supported by the ..... grants (Project No. .... and No. ....).

S. Salamat Ullah, and S. C. Liew are with the Department of Information Engineering, The Chinese University of Hong Kong, Hong Kong. Email: {ssullah, soung}@ie.cuhk.edu.hk. Gianluigi Liva is with the Institute of Communication and Networking, German Aerospace Center (DLR), Germany. Email: Gianluigi.Liva@dlr.de.

---

[1]Note that the decoder does not need to be modified with respect to the classical point-to-point setting

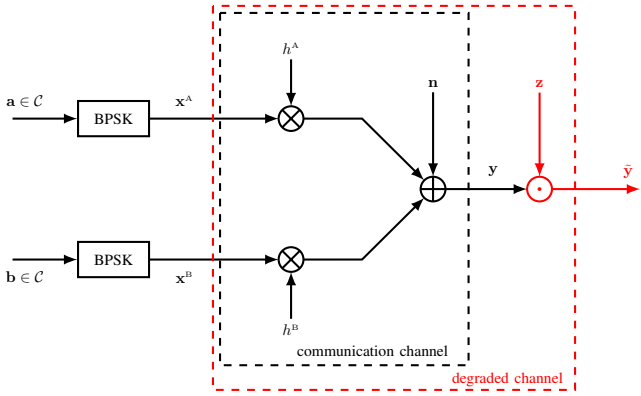

Fig. 1. Actual communication channel model and degraded channel model of uplink transmission in a two-way relay setting.

We focus our attention on the uplink phase. In particular, we assume the two users transmit simultaneously on the same frequency band using an $(N, K)$ binary linear block code $\mathcal{C}$. We further assume that the two users employ the binary phase shift keying (BPSK) modulation. The two transmissions are symbol-synchronous, i.e., at the receiver side, the symbols of the two users are aligned[2]. The two codewords transmitted by users $A$ and $B$ are written as $\mathbf{a}$ and $\mathbf{b}$. Denote the respective BPSK modulated versions by $\mathbf{x}^A$ and $\mathbf{x}^B$. The relay observes

$$\mathbf{y} = h^A \mathbf{x}^A + h^B \mathbf{x}^B + \mathbf{n}$$

where $h^A$, $h^B$ are the two (independent) complex channel coefficients, and $\mathbf{n} = (n_1, n_2, \ldots, n_N)$ is the additive white Gaussian noise (AWGN) contribution, with $n_i$, $i = 1, \ldots, N$, modeled as independent and identically distributed (i.i.d.) complex Gaussian random variables with zero mean and variance $2\sigma^2$. Figure 1 summarizes the reference model. The relay aims at decoding a binary linear combination (i.e., the XOR) codeword $\mathbf{c} = \mathbf{a} + \mathbf{b}$. When ML PNC decoding is employed, the relay computes

$$\hat{\mathbf{c}}_{\text{ML}} = \arg\max_{\mathbf{c} \in \mathcal{C}} \phi_{\text{ML}}(\mathbf{y}, \mathbf{c}) \quad (1)$$

where we define the *ML decoding metric* as

$$\phi_{\text{ML}}(\mathbf{y}, \mathbf{c}) := \sum_{\substack{\mathbf{a}, \mathbf{b}:\ \mathbf{a}+\mathbf{b}=\mathbf{c} \\ \mathbf{a}, \mathbf{b} \in \mathcal{C}}} p_{\mathbf{Y}|\mathbf{A}\mathbf{B}}(\mathbf{y}|\mathbf{a}, \mathbf{b}). \quad (2)$$

In (2), the channel transition probability density is

$$p_{\mathbf{Y}|\mathbf{A}\mathbf{B}}(\mathbf{y}|\mathbf{a}, \mathbf{b}) = \prod_{i=1}^{N} p_{Y|AB}(y_i|a_i, b_i)$$

with

$$p_{Y|AB}(y|a,b) := \frac{1}{2\pi\sigma^2} \exp\left(\frac{-|y - h^A \mathtt{m}(a) - h^B \mathtt{m}(b)|^2}{2\sigma^2}\right). \quad (3)$$

Here $\mathtt{m}(\cdot)$ is the modulation operation, with $\mathtt{m}(0) = +1$ and $\mathtt{m}(1) = -1$. From the equations above, we may restate (1) as

$$\hat{\mathbf{c}}_{\text{ML}} = \arg\max_{\mathbf{c} \in \mathcal{C}} \left[ \sum_{\substack{\mathbf{a}, \mathbf{b}:\ \mathbf{a}+\mathbf{b}=\mathbf{c} \\ \mathbf{a}, \mathbf{b} \in \mathcal{C}}} \left( \prod_{i=1}^{N} p_{Y|AB}(y_i|a_i, b_i) \right) \right].$$

[2]Milder conditions on symbol synchronism have been analyzed, among others, in [11]–[14].

In practical scenarios, sub-optimal PNC channel decoders such as XOR-CD are often used [6]. When dealing with XOR-CD, the demodulation and channel decoding tasks are separated. The (soft) demodulator provides the decoder with the bit-wise soft estimate

$$\lambda(y_i, c_i) := \sum_{\substack{a_i, b_i:\ a_i + b_i = c_i}} p_{Y|AB}(y_i|a_i, b_i).$$

Channel decoding then takes place in the same way as for a point-to-point channel, i.e., any off-the-shelf binary decoder can be employed. We introduce the reference decoder which computes

$$\hat{\mathbf{c}}_{\text{XC}} = \arg\max_{\mathbf{c} \in \mathcal{C}} \phi_{\text{XC}}(\mathbf{y}, \mathbf{c}), \quad (4)$$

where we define the decoding metric

$$\phi_{\text{XC}}(\mathbf{y}, \mathbf{c}) := \prod_{i=1}^{N} \lambda(y_i, c_i) \quad (5)$$

$$= \prod_{i=1}^{N} \sum_{\substack{a_i, b_i:\ a_i + b_i = c_i}} p_{Y|AB}(y_i|a_i, b_i) \quad (6)$$

$$= \sum_{\substack{\mathbf{a}, \mathbf{b}:\ \mathbf{a}+\mathbf{b}=\mathbf{c} \\ \mathbf{a}, \mathbf{b} \in \mathbb{F}_2^N}} p_{\mathbf{Y}|\mathbf{A}\mathbf{B}}(\mathbf{y}|\mathbf{a}, \mathbf{b}). \quad (7)$$

We refer to the XOR-CD scheme using the metric (5) as *maximum metric (MM) XOR-CD decoder*[3].

Let $P_{\text{B}}^{\text{XC}}$ be the block error probability under XOR-CD and $P_{\text{B}}^{\text{ML}}$ be the block error probability under ML PNC decoding. Since the ML decoder is optimal, $P_B^{\text{XC}} \geq P_B^{\text{ML}}$.

Comparing (2) with (5), we observe that the ML decoding metric does not admit a trivial factorization, whereas the decoding metric under an XOR-CD is given by the product of $N$ factors (one per observation). Observe also that the implementation of an XOR-CD scheme according to the rule (4) still entails in general a complexity growing exponentially with the block length. However, thanks to factorization of (5), one may use decoding algorithms which require input as bit-wise (or symbol-wise) metrics only; one can then also find fast algorithms to (4) that yield good solutions. This is not the case with ML PNC decoder, where the factorization of the ML decoding metric is not available. Considering, for example, a convolutional code (which has decoding complexity increasing linearly with the block length), the exact solution can be found for (4); for certain other codes such as LDPC code, decoding based on belief propagation (BP) (which generally converges fast) can be used to obtain good approximate solutions to (4).

## III. EQUIVALENT DEGRADED CHANNEL MODEL CONSTRUCTION

In this section, we introduce a (physically) degraded channel model that simplifies the calculation of the random coding error exponent for a PNC system employing XOR-CD. We first construct the degraded channel model and provide the ML PNC channel decoding metric for the degraded channel.

[3]Note that in (7) $\mathbf{a}$ and $\mathbf{b}$ are arbitrary vectors in $\mathbb{F}_2^N$ (i.e., they are not necessarily codewords), whereas in (2) the sum runs over all codeword pairs $\mathbf{a}, \mathbf{b}$ such that $\mathbf{a} + \mathbf{b} = \mathbf{c}$. The MM XOR-CD decoder is hence obtained as a relaxation of the optimum ML decoder.

We then show that ML decision of over the degraded channel is equal to the one provided by MM XOR-CD over the original channel. We refer to the proposed degraded channel as EDC.

To construct EDC, we modify the original channel by appending to the channel output a block which performs the multiplication of each channel output by a coefficient picked uniformly at random in $\{-1, +1\}$ (Figure 1). We denote by $\tilde{\mathbf{y}}$ the modified channel output. We have

$$\tilde{\mathbf{y}} = \mathbf{y} \odot \mathbf{z},$$

where $\odot$ denotes the Hadamard product and $\mathbf{z} = (z_1, \ldots, z_N)$ with elements modeled as i.i.d. Bernoulli random variables with $P_Z(-1) = P_Z(1) = 1/2$. Accordingly, we have

$$P_{\tilde{Y}|Y}(\tilde{y}|y) = \begin{cases} 1/2 & \text{if } \tilde{y} = +y \\ 1/2 & \text{if } \tilde{y} = -y \end{cases}$$

When ML PNC decoding is employed for EDC, the relay computes

$$\hat{\mathbf{c}}_{\text{MLD}} = \arg\max_{\mathbf{c} \in \mathcal{C}} \phi_{\text{ML}}^{\text{EDC}}(\tilde{\mathbf{y}}, \mathbf{c}),$$

where the ML PNC channel decoding metric for EDC is defined as

$$\phi_{\text{ML}}^{\text{EDC}}(\tilde{\mathbf{y}}, \mathbf{c}) := \sum_{\substack{\mathbf{a},\mathbf{b}:\ \mathbf{a}+\mathbf{b}=\mathbf{c} \\ \mathbf{a},\mathbf{b} \in \mathcal{C}}} p_{\tilde{\mathbf{Y}}|\mathbf{AB}}(\tilde{\mathbf{y}}|\mathbf{a},\mathbf{b}) \quad (8)$$

$$= \sum_{\substack{\mathbf{a},\mathbf{b}:\ \mathbf{a}+\mathbf{b}=\mathbf{c} \\ \mathbf{a},\mathbf{b} \in \mathcal{C}}} \prod_{i=1}^{N} p_{\tilde{Y}|AB}(\tilde{y}_i|a_i,b_i) \quad (9)$$

with the channel transition probability density

$$p_{\tilde{Y}|AB}(\tilde{y}_i|a_i,b_i) = \sum_{y_i = \pm \tilde{y}_i} p_{Y|AB}(y_i|a_i,b_i) p_{\tilde{Y}|Y}(\tilde{y}_i|y_i)$$

$$= \frac{1}{2}\left[p_{Y|AB}(\tilde{y}_i|a_i,b_i) + p_{Y|AB}(-\tilde{y}_i|a_i,b_i)\right]. \quad (10)$$

We now proceed to show that the ML PNC channel decoding on EDC is equivalent to XOR-CD on the original (non-degraded) channel. The following two lemmas will be useful for the purpose.

**Lemma 1.** *For the PNC system under consideration, we have*

$$p_{Y|AB}(y|a,b) = p_{Y|AB}(-y|\bar{a},\bar{b}),$$

*where $\bar{a}$ ($\bar{b}$) is the binary complement of $a$ ($b$).*

*Proof.* Due to the adoption of BPSK modulation at the end users, we have $\mathtt{m}(a) = (-1)^a$ and $\mathtt{m}(b) = (-1)^b$. From (3), we have

$$p_{Y|AB}(-y|\bar{a},\bar{b}) = \frac{1}{2\pi\sigma^2}\exp\left(\frac{-|-y-h^{\text{A}}\mathtt{m}(\bar{a})-h^{\text{B}}\mathtt{m}(\bar{b})|^2}{2\sigma^2}\right)$$

$$= \frac{1}{2\pi\sigma^2}\exp\left(\frac{-|-y-h^{\text{A}}(-1)^{a+1}-h^{\text{B}}(-1)^{b+1}|^2}{2\sigma^2}\right)$$

$$= \frac{1}{2\pi\sigma^2}\exp\left(\frac{-|-y+h^{\text{A}}(-1)^{a}+h^{\text{B}}(-1)^{b}|^2}{2\sigma^2}\right)$$

$$= \frac{1}{2\pi\sigma^2}\exp\left(\frac{-|y-h^{\text{A}}(-1)^{a}-h^{\text{B}}(-1)^{b}|^2}{2\sigma^2}\right)$$

$$= p_{Y|AB}(y|a,b).$$

$\square$

Lemma 1 formalizes the symmetry in the superimposed constellation at the relay, which is obvious from observing Figure 2. We next present Lemma 2 that states the function given therein is an even function of its associated observation.

**Lemma 2.** *For the PNC system under consideration, the summation of the channel transition probability of the original (non-degraded) channel over the XOR coded-bit is an even function of the observation variable $Y$, i.e.,*

$$\sum_{a_i,b_i:\ a_i+b_i=c_i} p_{Y|AB}(y_i|a_i,b_i) = \sum_{a_i,b_i:\ a_i+b_i=c_i} p_{Y|AB}(-y_i|a_i,b_i).$$

*Proof.* Take left hand side of the above equation, expand the summation and then apply Lemma 1 to get

$$\sum_{a_i,b_i:\ a_i+b_i=c_i} p_{Y|AB}(y_i|a_i,b_i)$$

$$= p_{Y|AB}(y_i|a_i,a_i+c_i) + p_{Y|AB}(y_i|\bar{a}_i,\bar{a}_i+c_i)$$

$$= p_{Y|AB}(-y_i|\bar{a}_i,\bar{a}_i+c_i) + p_{Y|AB}(-y_i|a_i,a_i+c_i)$$

$$= \sum_{a_i,b_i:\ a_i+b_i=c_i} p_{Y|AB}(-y_i|a_i,b_i).$$

$\square$

We next show that the ML PNC channel decoder operating on EDC outputs a decision equal to that of XOR-CD over the original (non-degraded) channel.

**Theorem 1.** *The ML decision over the EDC coincides with the MM XOR-CD decision over the (original) communication channel.*

*Proof.* To prove the theorem, we have to show that the ML PNC channel decoding metric for EDC is equal to (or some scalar multiple of) the decoding metric of XOR-CD for the (original) communication channel, i.e.,

$$\phi_{\text{ML}}^{\text{EDC}}(\tilde{\mathbf{y}},\mathbf{c}) \propto \phi_{\text{XC}}(\mathbf{y},\mathbf{c}),$$

where $\phi_{\text{ML}}^{\text{EDC}}(\tilde{\mathbf{y}},\mathbf{c})$ is defined in (8) and $\phi_{\text{XC}}(\mathbf{y},\mathbf{c})$ is defined in (5).

After applying Lemma 1, i.e., $p_{Y|AB}(-\tilde{y}_i|a_i,b_i) = p_{Y|AB}(\tilde{y}_i|\bar{a}_i,\bar{b}_i)$, to (10) and then putting it in (9), we get

$$\phi_{\text{ML}}^{\text{EDC}}(\tilde{\mathbf{y}},\mathbf{c}) = \sum_{\substack{\mathbf{a},\mathbf{b}:\ \mathbf{a}+\mathbf{b}=\mathbf{c} \\ \mathbf{a},\mathbf{b} \in \mathcal{C}}} \prod_{i=1}^{N} \frac{1}{2}\left[p_{Y|AB}(\tilde{y}_i|a_i,b_i) + p_{Y|AB}(\tilde{y}_i|\bar{a}_i,\bar{b}_i)\right].$$

Observe that $a_i$ and $b_i$ in the equation above turn out to be dummy variables. Their individual values do not matter; only their sum $c_i$ matters. So, for a given $\mathbf{c}$, we have

$$\phi_{\text{ML}}^{\text{EDC}}(\tilde{\mathbf{y}},\mathbf{c}) \propto \prod_{i=1}^{N} \sum_{a_i,b_i:\ a_i+b_i=c_i} p_{Y|AB}(\tilde{y}_i|a_i,b_i),$$

$$= \prod_{i=1}^{N} \sum_{a_i,b_i:\ a_i+b_i=c_i} p_{Y|AB}(y_i|a_i,b_i)$$

$$= \phi_{\text{XC}}(\mathbf{y},\mathbf{c}),$$

where the first equality follows from Lemma 2 since $\tilde{y}_i = \pm y_i$ and the summation term is an even function of $y_i$. The second equality follows from (6).

$\square$

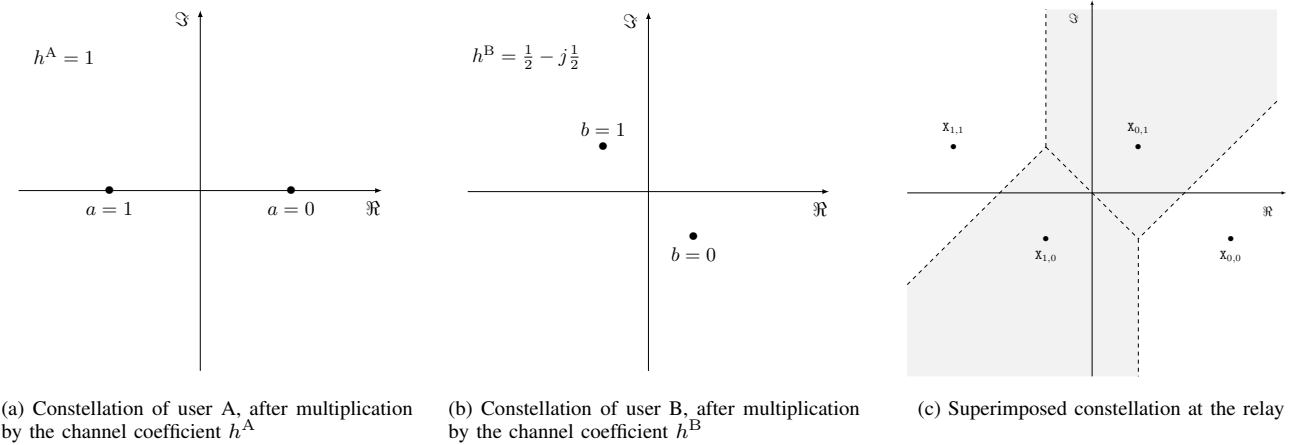

Fig. 2. Constellations of users A and B, and superimposed constellation at the relay. The superimposed constellation point associated with the transmission on the couple $(a, b)$ is denoted by $\mathtt{x}_{a,b}$.

**Observation 1.** *From the above theorem, we see that the performance of XOR-CD in the original (non-degraded) channel can be fully characterized by analyzing the transmission with a linear block code $\mathcal{C}$ over the (virtual) memory-less point-to-point channel*

$$p_{\tilde{\mathbf{Y}}|\mathbf{C}}(\tilde{\mathbf{y}}|\mathbf{c}) = \prod_{i=1}^{N} p_{\tilde{Y}|C}(\tilde{y}_i|c_i)$$

*with the transition probability density*

$$p_{\tilde{Y}|C}(\tilde{y}_i|c_i) = p_{\tilde{Y}|C}(y_i|c_i) = \frac{1}{2}\left[\sum_{\substack{a_i,b_i:\\a_i+b_i=c_i}} p_{Y|AB}(y_i|a_i,b_i)\right]. \quad (17)$$

## IV. RANDOM CODING ERROR EXPONENTS

Given Observation 1, the derivation of the random coding error exponent under XOR-CD follows simply by deriving the error exponent for the virtual point-to-point channel with transition probability density given by (17). Recall that Gallager's random coding bound (RCB) [10] on the average block error probability $\bar{P}_{\mathrm{B}}$ of random $(N,K)$ codes has the form

$$\bar{P}_{\mathrm{B}} \leq 2^{-N E_G(R)}, \quad (18)$$

where $N$ is a block length, $R = K/N$ is the code rate and $E_G(R)$ is the random coding error exponent. The bound (18) holds also for the ensemble of linear random codes (see e.g. [15]), and thus applies to the XOR-CD PNC setting.

We denote in the following the random variable associated with $\tilde{y}$ by $\tilde{Y}$, and the two independent and uniformly distributed binary random variables associated with XOR-coded bits by $C$, $C'$. Under perfect CSI, the random coding error exponent is

$$E_G(R) = \max_{0 \leq \rho \leq 1}\left[E_0(\rho) - \rho R\right]$$

with

$$E_0(\rho) := -\log_2 \mathbb{E}\left[\left(\frac{\mathbb{E}\left[p_{\tilde{Y}|C}(Y|C')^{\frac{1}{1+\rho}}\big|Y\right]}{p_{\tilde{Y}|C}(Y|C)^{\frac{1}{1+\rho}}}\right)^{\rho}\right].$$

Observe that, remarkably, the random coding error exponent under XOR-CD, given above, for linear random codes is exactly the same as the random coding error exponent under ML PNC channel decoding, given in [9], for linear codes.

Under mismatched CSI, the relay does not possess perfect knowledge of the channel coefficients but rather two estimates $\hat{h}^{\mathrm{A}}, \hat{h}^{\mathrm{B}}$. The decoder hence operates with the mismatched metric

$$q(y,c;\hat{h}^{\mathrm{A}},\hat{h}^{\mathrm{B}}) = \sum_{a,b:\ a+b=c} \exp\left(\frac{-|y - \hat{h}^{\mathrm{A}}\mathtt{m}(a) - \hat{h}^{\mathrm{B}}\mathtt{m}(b)|^2}{2\sigma^2}\right).$$

The random coding error exponent is in this case [16]–[18]

$$E_G(R) = \max_{0 \leq \rho \leq 1}\sup_{s \geq 0}\left[E_0(\rho,s) - \rho R\right]$$

where

$$E_0(\rho,s) := -\log_2 \mathbb{E}\left[\left(\frac{\mathbb{E}\left[q(Y,C';\hat{h}^{\mathrm{A}},\hat{h}^{\mathrm{B}})^s\big|Y\right]}{q(Y,C;\hat{h}^{\mathrm{A}},\hat{h}^{\mathrm{B}})^s}\right)^{\rho}\right],$$

where the inner expectation is with respect to the random variable $C'$ and outer expectation is with respect to the random variables $Y$ and $C$.

### A. Example of Application

Finite-length performance benchmarks (see e.g. [19]–[21] for short and moderate-length channel codes are gaining much interest, following emerging applications which entail transmissions of short messages [22]. In this context, bounds as the one provided by (18) are particularly useful to understand the quality of error correcting schemes used in practical implementations. As an example of application, let us compare the performance of a low-density parity-check (LDPC) coded PNC scheme with the random coding bound. The comparison is even more insightful for the mismatched CSI case, considering the limitations in the precision of the channel estimation when short blocks are transmitted [23].

Consider the case where a $(128, 64)$ irregular repeat-accumulate (IRA) code [24] is used to encode the information at the end nodes. For the information part, a regular variable node degree of $4$ is used. We assume the signals of the two end users are phase aligned at the relay and that $h^\text{A} = h^\text{B} = 1$. Both users employ BPSK modulation. For the mismatched setting, we assume that the relay node has mismatched CSI. We provide two examples of mismatched CSI, i.e.

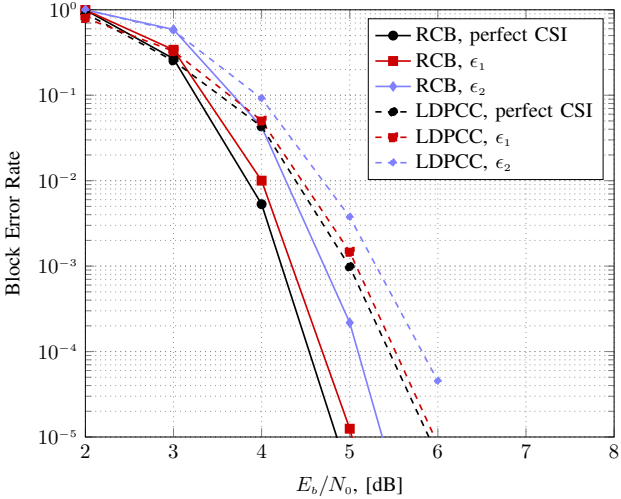

Fig. 3. Block error rate vs. $E_b/N_0$ for a $(128, 64)$ LDPC-coded PNC under perfect CSI and mismatched CSI, compared with the RCB for XOR-CD.

$$\epsilon_1 := \hat{h}^\text{A} - h^\text{A} = \hat{h}^\text{B} - h^\text{B} = -0.1 - j0.1$$

and

$$\epsilon_2 := \hat{h}^\text{A} - h^\text{A} = \hat{h}^\text{B} - h^\text{B} = -0.2 - j0.2.$$

Albeit based on XOR-CD, the LDPC decoder does not compute (5) but resorts to the simpler BP decoder by providing at the input of the variable nodes log-likelihood ratios (LLRs) in the form

$$L_i = \ln\left(\frac{p_{\tilde{Y}|C}(y_i|0)}{p_{\tilde{Y}|C}(y_i|1)}\right)$$

for $i = 1, \ldots, N$.

Figure 3 shows the block error rate vs. $E_b/N_0$ (being $E_b$ the eneregy per information bit per user, and $N_0$ the single-sided noise power spectral density) for the designed code, together with the random coding bound under both perfect and imperfect (mismatched) CSI. The designed code, though not able to attain the upper bound on the random code performance, is capable of achieving a block error rate of $10^{-4}$ within 1 dB from the corresponding random coding bound, consistently with what observed for similar code constructions on the binary-input AWGN channel [25].

## V. Conclusion

This work derives the random coding error exponent for the uplink phase of a TWRC operated with PNC, where the two end users employ the same binary linear block code. Unlike prior work, this work derives the error exponent for the sub-optimum but more practical XOR-CD approach. Remarkably, the error exponent under optimum ML decoding can be achieved even under the simpler XOR-CD strategy. The derivation is based on the introduction of an equivalent degraded channel model which simplifies the XOR-CD performance analysis.